\documentclass[twocolumn,rmp,showpacs]{revtex4}%
\usepackage{amsfonts}
\usepackage{amsmath}
\usepackage{amssymb}
\usepackage{graphicx}
\usepackage{times}%
\providecommand{\U}[1]{\protect\rule{.1in}{.1in}}

\begin{document}
\title{Broken spin-Hall accumulation symmetry by magnetic field \\and coexisted Rashba and Dresselhaus interactions}
\author{Son-Hsien Chen}
\email{d92222006@ntu.edu.tw}
\affiliation{Department of Physics, National Taiwan University, Taipei 10617, Taiwan}
\author{Ming-Hao Liu}
\affiliation{Department of Physics, National Taiwan University, Taipei 10617, Taiwan}
\author{Kuo-Wei Chen}
\affiliation{Department of Physics, National Taiwan University, Taipei 10617, Taiwan}
\author{Ching-Ray Chang}
\affiliation{Department of Physics, National Taiwan University, Taipei 10617, Taiwan}

\pacs{73.21.-b, 71.70.Ej, 75.60.Ej}
\keywords{spin-Hall, spin accumulation, spin-orbit interaction, Rashba}
\begin{abstract}
The spin-Hall effect in the two-dimensional electron gas (2DEG) generates
symmetric out-of-plane spin $S_{z}$ accumulation about the current axis, in
the absence of external magnetic field. Here we employ the real space
Landauer-Keldysh formalism\cite{Kel} by considering a four-terminal setup to
investigate the circumstances in which this symmetry is broken. For absence of
Dresselhaus interaction, starting from the applied out-of-plane $B$
corresponding to Zeeman splitting energy $\Delta=0$ to $0.5$ times the Rashba
hopping energy $t_{SO}^{R}$, the breaking process is clearly seen. The
influence of the Rashba interaction on the magnetization of the 2DEG is
studied herein. For coexisted Rashba $t_{SO}^{R}$ and Dresselhaus $t_{SO}^{D}$
spin-orbit couplings in the absence of $B$, interchanging $t_{SO}^{R}$ and
$t_{SO}^{D}$ reverses the entire accumulation pattern.

\end{abstract}
\maketitle

The generation and transport of spin currents dominates the applications of
spintronics. The spin-orbit (SO) interaction, which couples the electric
degree of freedom with the magnetic one serves as\emph{ }the mechanism to
achieve this. A number of basic designs$^{\text{\cite{design}}}$ for
spintronic devices, such as field-effect switches, spin transistors, spin
filters, and spin waveguides, have been proposed by taking advantage of this
interaction to control spins. One of the phenomena originating from the SO
interaction is the spin-Hall (SH) effect$^{\text{\cite{Sinova}}}$ in which a
transverse spin current is induced by a longitudinal electrical current. The
semiclassical SO force$^{\text{\cite{SOforce}}}$ proportional to
$(\mathbf{p}\times\mathbf{e}_{z})\sigma$, oppositely deflects the spin up
($\sigma=1$) and down ($\sigma=-1$) wave packets with momentum $\mathbf{p}$ in
the transverse directions so that different spins accumulate in the lateral
edges. The effect is particularly notable in that it produces a spin current
with no magnetic field applied, and no accompanying charge current present.
Recent experimental work using scanning Kerr
microscopy$^{\text{\cite{experiment1,experiment2,experiment3,experiment4}}}$
in \textit{n}-type unstrained GaAs, strained InGaAs$^{\text{\cite{experiment1}%
}}$ and 2DEG,$^{\text{\cite{experiment3}}}$ have inspired a host of
theoretical studies on SH effect.

The 2DEG confined in the InGaAs/InAlAs semiconductor
heterostructure$^{\text{\cite{heture}}}$ possesses intrinsic SO coupling known
as the Rashba and Dresselhaus interactions. The inversion asymmetry of the
structure gives this system the Rashba interaction with adjustable strength
via the gate voltage$^{\text{\cite{heture,Das}}}$ while the bulk inversion
asymmetry$^{\text{\cite{Dresselhaus}}}$ gives rise to the Dresselhaus
interaction whose strength is material dependent. A number of studies
regarding the intrinsic SH (ISH) effect in the 2DEG have been reported, but
most of these focus on evaluating the spin current which is, due to its
nonconservation, not easily measured. The SO coupling leads an electron to
precess around a momentum-dependent effective magnetic field which produces a
source or a sink of the spin current in the continuity equation. Due to this
nonconservation the spin current is not uniquely
defined.$^{\text{\cite{notuni1,notuni2}}}$ Although a possible definition of a
conserved spin current has been suggested in Ref. \onlinecite{condef}, the
present work follows Ref. \onlinecite{Kel} by employing the Landauer-Keldysh
formalism in the four-terminal setup to investigate a more directly measurable
physical quantity: the accumulation of the out-of-plane spin $S_{z}$. If only
the Rashba interaction exists then the accumulation induced by the electric
potential deposits spins symmetrically along the transverse direction, i.e.,
the up-spins accumulate on one side while the down-spins accumulate
symmetrically on the other. The majority of current studies focuses on how
this SH symmetry (SHS) is generated. By contrast, in the paper we answer the
essential question of the circumstances under which the SHS is broken.

\begin{figure}
[ptb]
\begin{center}
\includegraphics[
trim=0.806850in -0.010149in 0.040954in 0.017480in,
height=3.1808in,
width=2.9767in
]%
{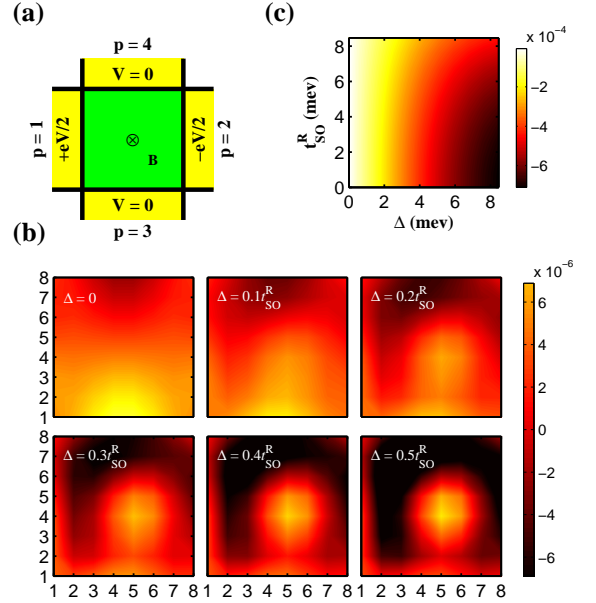}%
\caption{(Color online) (a) Illustration of the Landauer four-terminal setup
with the out-of-plane $B$. Contacting the 2DEG (green square) with four ideal
semi-infinite leads (yellow rectangles) $p=1$ (left), $2$ (right), $3$
(bottom), $4$ (top) which are biased with $eV_{1}=-eV_{2}=eV/2$,
$eV_{3}=eV_{4}=0$ and $eV=10^{-3}t_{0},$ the spatial out-of-plane $S_{z}$
accumulations, in units of $\hbar/2$ with Zeeman splitting $\Delta$ being
varied from $0$ to $0.5t_{SO}^{R}$, are plotted in (b)$.$ The magnetization,
obtained by summing over every accumulation on each site, in units of
$\hbar/2$, decreases with $t_{SO}^{R}$ but increases with $\Delta$ as shown in
(c).}%
\label{fig1}%
\end{center}
\end{figure}

With the help of the finite difference method,$^{\text{\cite{Dattabook}}}$ the
linear Rashba and Dresselhaus model with applied field $B$ is, in the tight
binding limit, expressed as%

\begin{align}
\hat{H}^{\left(  c\right)  }  &  =\sum_{\mathbf{m\sigma}}\varepsilon
_{\mathbf{m}}c_{\mathbf{m\sigma}}^{\dagger}c_{\mathbf{m\sigma}}\label{H1}\\
&  +\sum_{\mathbf{mm}^{\prime}\sigma\sigma^{\prime}}c_{\mathbf{m}\sigma
}^{\dagger}t_{\mathbf{mm}^{\prime}}^{\sigma\sigma^{\prime}}c_{\mathbf{m}%
^{\prime}\sigma^{\prime}}+\sum_{\mathbf{m\sigma}}\sigma\frac{\Delta}%
{2}c_{\mathbf{m\sigma}}^{\dagger}c_{\mathbf{m\sigma}},\nonumber
\end{align}
where $c_{\mathbf{m}\sigma}^{\dagger}$ ($c_{\mathbf{m}^{\prime}\sigma^{\prime
}}$) denotes the creation (annihilation) operator for $\sigma=+1$ (spin-up) or
$\sigma=-1$ (spin-down) at the site $\mathbf{m}=(\mathbf{m}_{x},\mathbf{m}%
_{y})$. The electric potential and disorder can be accounted for by the
on-site energy $\varepsilon_{\mathbf{m}}$. To see how SHS is destroyed either
by $B$ or coexisted Dresselhaus and Rashba interactions, we consider the clean
limit and set the tight-binding bottom energy, for convenience, to be
$-4t_{0}$ (corresponding to $\varepsilon_{\mathbf{m}}=0$), with $t_{0}%
=\hbar^{2}/2ma^{2}$ being the hopping energy. The Zeeman splitting
$\Delta=-e\hbar B/mc$ is induced by the magnetic field while the Rashba
(Dresselhaus) SO coupling $t_{SO}^{R(D)}=\alpha(\beta)/2a$ is taken into
account by the neareast-neighbor hopping matrix element $t_{\mathbf{mm}%
^{\prime}}^{\sigma\sigma^{\prime}}=\left\langle \sigma\right\vert -t_{0}%
I_{s}-it_{SO}^{R}\sigma_{y}-it_{SO}^{D}\sigma_{x}|\sigma^{^{\prime}}\rangle$
for $\mathbf{m}=\mathbf{m}^{\prime}+\mathbf{e}_{x}$, and $\left\langle
\sigma\right\vert -t_{0}I_{s}+it_{SO}^{R}\sigma_{x}+it_{SO}^{D}\sigma
_{y}|\sigma^{^{\prime}}\rangle$ for $\mathbf{m}=\mathbf{m}^{\prime}%
+\mathbf{e}_{y}$. Here the identity matrix in spin space is $I_{s}$, the
lattice constant is $a$, and the Rashba (Dresselhaus) coupling strength is
$\alpha$ ($\beta$). In order to subject the conductor to an electric potential
difference along the $x$ axis we consider the Landauer setup with the four
ideal leads $p=1$ (left)$,$ $2$ (right)$,$ $3$ (bottom)$,4$ (top) shown in
Fig. \ref{fig1}(a).

To acquaint the reader with the method employed herein we present here a brief
review of the Landauer-Keldysh Formalism. In general, in a conductor the
non-equilibrium spin accumulation at time $t$, $\left\langle S_{\mathbf{m}%
}^{z}(t)\right\rangle $ $=\hbar/2\sum_{\sigma}\sigma\left\langle
c_{\mathbf{m\sigma}}^{\dagger}(t)c_{\mathbf{m\sigma}}(t)\right\rangle
=\hbar/2\sum_{\sigma}\sigma\left[  \left\langle \mathbf{m}\sigma\right\vert
\mathbf{G}^{<(c)}(t,t)\left\vert \mathbf{m}\sigma\right\rangle \hbar/i\right]
,$ depends on the switching time $t_{s}$ (at which leads are brought into
contact) via the lesser Green function
\begin{align}
&  \mathbf{G}^{<\left(  c\right)  }(t_{1},t_{1}^{^{\prime}})\label{leserg1}\\
&  =\int_{t_{s}}^{\infty}dt_{2}\int_{t_{s}}^{\infty}dt_{3}\mathbf{G}^{r\left(
c\right)  }(t_{1},t_{2})\mathbf{\Sigma}^{<\left(  c\right)  }(t_{2,}%
t_{3})\mathbf{G}^{r\left(  c\right)  \dagger}(t_{3},t_{1}^{^{\prime}%
})\nonumber\\
&  +\mathbf{G}^{r\left(  c\right)  }(t_{1},t_{s})\mathbf{G}^{<\left(
c\right)  }(t_{s},t_{s})\mathbf{G}^{r\left(  c\right)  \dag}(t_{s}%
,t_{1}^{\prime}),\nonumber
\end{align}
which includes both the steady state (second line), and also the transient
state (third line). For the measuring time $t$ much later than $t_{s}$ (this
is the case of interest here) one can approximately write in Eq.
(\ref{leserg1}) $t_{s}=-\infty$. The transient solution can then be neglected,
and $\mathbf{G}^{<(c)}(t_{1},t_{1}^{^{\prime}})$ $=\mathbf{G}^{<(c)}(\tau)$
depends only on the the time interval $\tau\equiv t_{1}-t_{1}^{\prime}.$ A
Fourier transformation of Eq. (\ref{leserg1}) then yields the kinetic
equation
\begin{equation}
\mathbf{G}^{<(c)}(E)=\mathbf{G}^{r(c)}(E)\mathbf{\Sigma}^{<(c)}(E)\mathbf{G}%
^{r(c)\dagger}(E) \label{Kinetic}%
\end{equation}
with the retarded Green function $\mathbf{G}^{r(c)}(E)=[E-\mathbf{H}%
^{(c)}-\mathbf{\Sigma}^{(c)}(E)]^{-1}$, so that the steady accumulation is
expressed as $S_{\mathbf{m}}^{z}(t=0)=\frac{\hbar}{2}\sum_{\sigma
\sigma^{\prime}=\pm1}\left\langle \sigma\right\vert \mathbf{\sigma}_{z}%
|\sigma^{^{\prime}}\rangle\int_{-\infty}^{\infty}dE(\mathbf{G}(E)^{<(c)}%
)_{\mathbf{mm}\sigma^{\prime}\sigma}/2\pi i$. The lead interacts with the
conductor through the self-energy$^{\text{\cite{Dattabook}}}$ $\mathbf{\Sigma
}^{(c)}=\sum_{p}\mathbf{\Sigma}_{p}^{(c)},$ with matrix elements $\left\langle
\mathbf{m,}\sigma\right\vert \mathbf{\Sigma}_{p}^{(c)}|\mathbf{m}^{\prime
},\sigma^{^{\prime}}\rangle\equiv\Sigma_{p\mathbf{m},\mathbf{m}^{\prime}%
}^{(c)\sigma\sigma^{\prime}}=t_{0}^{2}g^{(p)}(\mathbf{r}_{\mathbf{m}%
},\mathbf{r}_{\mathbf{m}^{\prime}}^{\prime})\delta_{\sigma\sigma^{\prime}}$
for $\mathbf{m}$ and $\mathbf{m}^{\prime}$ (in the conductor) being adjacent
points to $\mathbf{r}_{\mathbf{m}}$ and $\mathbf{r}_{\mathbf{m}^{\prime}%
}^{\prime}$ (in leads), and $\Sigma_{p\mathbf{m},\mathbf{m}^{\prime}%
}^{(c)\sigma\sigma^{\prime}}=0$ otherwise. The hopping energy $t_{0}$ for
these points allows electrons to flow through the interfaces. While the lesser
self-energy is written as%
\begin{equation}
\mathbf{\Sigma}^{<(c)}(E)=-2i\operatorname{Im}\mathbf{\Sigma}_{p}%
(E-eV_{p})f(E-eV_{p}), \label{Lesser}%
\end{equation}
where $f(E)=1/\left[  1+\exp(-E/k_{B}T)\right]  $ is the Fermi-Dirac
distribution and the quasi-particle escaping time, in the conductor, is
inversely proportional to $\operatorname{Im}\mathbf{\Sigma}^{(c)}(E)$. The
retarded Green function $\hat{g}^{(p)}$ for the isolated lead $p$ can be
evaluated by considering the single particle Green operator $\hat{g}%
^{(p)}=1/(E+i0_{+}-\hat{H}^{(p)})$ in the eigenfunction expansion
$g^{(p)}(\mathbf{r}_{1},\mathbf{r}_{2})=\sum_{n\gamma}\psi_{n\gamma}%
^{(p)}(r_{1l},r_{1t})\psi_{n\gamma}^{(p)\ast}(r_{2l},r_{2t})/(E+i0_{+}%
-\varepsilon_{n\gamma}^{(p)})$, where $\mathbf{r=(}r_{l},$ $r_{t})$ is the
position vector (within the lead) composed by the longitudinal component
$r_{l}$ and the transverse component $r_{t}$ with $n$ and $\gamma$ accounting
the transverse and longitudinal modes, respectively. The eigenfunction
$\psi_{n\gamma}^{\left(  p\right)  }\left(  r_{l},r_{t}\right)  =2\sin(n\pi
r_{t}/W)\sin(\gamma\pi r_{l}/L)/(a\sqrt{WL})$, with the normalization
$\sum_{n\gamma}$ $\psi_{n\gamma}^{(p)}(\mathbf{r}^{\prime})\psi_{n\gamma
}^{(p)\ast}(\mathbf{r)}=\delta_{\mathbf{rr}^{\prime}}$, is obtained by solving
the eigenequation $\hat{H}_{p}\psi_{n\gamma}^{(p)}(r_{l},r_{t})=[-(\hbar
^{2}/2m)d^{2}/dr_{l}^{2}+V_{conf}^{(p)}(r_{t})]\psi_{n\gamma}^{(p)}%
(r_{l},r_{t})=\varepsilon_{n\gamma}\psi_{n\gamma}^{(p)}(r_{l},r_{t})$ under
the hard-wall boundary condition in which the confining potential
$V_{conf}^{(p)}$ $=$ $\infty$ ($0$) outside (inside) the lead $p$. The width
(length) of the lead is $W$ ($L$). For semi-infinite lead $L\rightarrow\infty
$, $g^{(p)}(\mathbf{r}_{1},\mathbf{r}_{2})$ can be directly computed by
replacing the summation of longitudinal modes $\sum_{\gamma}$ with $\int
_{0}^{\infty}d\gamma L/\pi.$

\begin{figure}
[ptb]
\begin{center}
\includegraphics[
trim=0.726467in 0.000000in 0.064214in 2.833273in,
height=1.5921in,
width=3.3537in
]%
{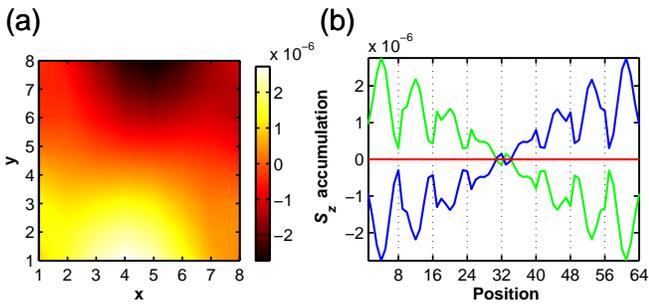}%
\caption{(Color online) (a) The spatial $S_{z}$ accumulation with $\left(
t_{SO}^{R},t_{SO}^{D}\right)  =\left(  0.1t_{0},0.05t_{0}\right)  .$ (b) The
$S_{z}$ accumulations in units of $\hbar/2$ as functions of positions with
parameters $\left(  t_{SO}^{R},t_{SO}^{D}\right)  =\left(  0.1t_{0}%
,0.05t_{0}\right)  $ (green) and $\left(  t_{SO}^{R},t_{SO}^{D}\right)
=\left(  0.05t_{0},0.1t_{0}\right)  $ (blue). The red line sums up these two
functions.}%
\label{fig2}%
\end{center}
\end{figure}

Consider now the InGaAs/InAlAs heterostructure$^{\text{\cite{heture}}}$ with
typical parameters. The effective mass $m=0.05m_{e}$ ($m_{e}$ is the electron
mass) and the lattice constant $a=3$ nm yield the hopping energy $t_{0}=84.68$
m$%
\operatorname{eV}%
$. We set $t_{SO}^{R}=0.1t_{0}$, $eV_{1}=-eV_{2}=eV/2$, $eV_{3}=eV_{4}=0$,
$eV=10^{-3}t_{0}$ and conductor size to be $8a\times8a$. Select the Fermi
energy $E_{F}=-3.8t$ close to the band bottom at $-4t_{0},$ so that the
tight-binding approximation valid. To examine how $\Delta$ and the coexistence
of $t_{SO}^{R}$ and $t_{SO}^{D}$ affect the SHS, we show the spatial $S_{z}$
accumulation in units of $\hbar/2$ in Fig. \ref{fig1} (with $t_{SO}^{D}=0$)
and Fig. \ref{fig2} (with $\Delta=0$).

Obviously, the applied field $B$ polarizes the 2DEG or, from perspective of
the band theory, it induces a Zeeman splitting $\Delta$ such that the SHS is
destroyed. Starting from $\Delta=0$ and going to $\Delta=0.5t_{SO}^{R},$ two
effects are found to break the symmetry [see Fig. \ref{fig1}(b)]:\emph{ }(i)
The area of the majority spins is enlarged. (ii) The magnitude of the majority
magnetization is increased. On the other hand, for parameters $\Delta$ and
$t_{SO}^{R}$ varying between $0$ and $0.1t_{0}$, the magnetization (or the
total $z$-polarization) of the system, obtained by summing over every $S_{z}$
on each site, is decreased with increasing Rashba interaction as shown in
[Fig. \ref{fig1}(c)]. To explain this, we note that, under the Rashba
interaction, electrons precess with respect to an in-plane ($x$-$y$) effective
Rashba magnetic field which yields vanishing mean $z$-polarization so that
magnetization is reduced.%

Even $B$ is absent, the SHS can also be broken by the coexistence of
$t_{SO}^{R}$ and $t_{SO}^{D}$. Consider the Dresselhaus coupling $t_{SO}%
^{D}=0.1t_{0}$ ($0.05t_{0}$), corresponding to $\beta\thickapprox
\gamma\left\langle k_{z}^{2}\right\rangle =0.5$ ($0.025$) $%
\operatorname{eV}%
\operatorname{\text{\AA}}%
$, assuming a $z$ direction quantum well width $22$ $%
\operatorname{\text{\AA}}%
$ ($31%
\operatorname{\text{\AA}}%
$) with typical value of the coefficient$^{\text{\cite{Knap}}}$ $\gamma=25$ $%
\operatorname{eV}%
\operatorname{\text{\AA}}%
^{3}$. Our results indicate that the SHS is preserved by the presence of
$t_{SO}^{R}$ or $t_{SO}^{D}$ alone, but it is destroyed if the Rashba and
Dresselhaus interactions coexist. Figure \ref{fig2}(a) shows the accumulation
pattern for $t_{SO}^{R}=0.1t_{0}$, $t_{SO}^{D}=0.05t_{0}$, and $B=0$. In
comparison to the $\Delta=0$ result of Fig. \ref{fig1}(b) it is tilted upwards
in the left hand region, and downwards in the right hand region. Furthermore,
if the strength of the Rashba $t_{SO}^{R}$ and the Dresselhaus $t_{SO}^{D}$
interactions is interchanged then the pattern is entirely reversed, i.e., the
up (down) accumulation becomes the down (up) accumulation. To illustrate this
effect we label the spin component $S_{z}$ by numbering the position of each
site row by row. For example, $\mathbf{m}=(a,a)$ is labeled by position $1$
and $\mathbf{m}=(8a,8a)$ is labeled by position $64$. The accumulation of the
component $S_{z}$ in the absence of the field $B$ is plotted in Fig.
\ref{fig2}(b) as a function of position for two special cases: (i) $\left(
t_{SO}^{R},t_{SO}^{D}\right)  =\left(  0.1t_{0},0.05t_{0}\right)  $ denoted by
the green line, and (ii) $\left(  t_{SO}^{R},t_{SO}^{D}\right)  =\left(
0.05t_{0},0.1t_{0}\right)  $ denoted by the blue line. Summing up these two
functions at every position we obtain zero accumulation everywhere (the red
line). This suggests an inverse correlation between the cases (i) and (ii). In
the case of equal strengths, $t_{SO}^{R}=t_{SO}^{D}$, one therefore expects
that $S_{z}$ accumulates nowhere, since swapping $t_{SO}^{R}$ and $t_{SO}^{D}$
has no effect. Finally, we recall that we address a finite 2DEG. Comparing our
results with previous works on infinite systems we identify the predicted sign
change$^{\text{\cite{Sign}}}$ in the spin-Hall conductivity.

In conclusion, we investigate the out-of-plane $S_{z}$ spin accumulation using
the Landauer-Keldysh formalism in a four terminal setup. Taking into account
the Rashba $t_{SO}^{R}$ and the Dresselhaus $t_{SO}^{D}$ couplings and Zeeman
splitting $\Delta$ we obtain an accumulation pattern different from the one
which results from pure Rashba interactions. In particular, destructions of
the SHS are found in two special cases: in the presence of a magnetic field
$B$ and in the presence of coexisting Rashba and Dresselhaus couplings. In the
former case, beginning with both $\Delta=0$ and $t_{SO}^{D}=0$, the applied
field $B$ breaks the SHS by not only extending the area of the majority spin
accumulation, but also by strengthening its magnitude. Meanwhile, the
gate-voltage-tunable Rashba interaction reduces the magnetization induced by
the Zeeman splitting. In the latter case (where both SO couplings are
present), interchanging the Rashba and Dresselhaus interactions reverses the
whole accumulation pattern. These features thus provide an electric control of
a magnetic property.

This work is supported by the Republic of China National Science Council Grant
No. 95-2112-M-002-044-MY3.

\end{document}